# An integrated approach to doped thin films with strain tunable magnetic anisotropy: Powder synthesis, target preparation and pulsed laser deposition of Bi:YIG

Pathikumar Sellappan[1,2], Chi Tang[3], Jing Shi[1,3] and Javier E. Garay[1,2] *

[1] Materials Science & Engineering Program, University of California, Riverside, CA, 92521, USA

[2] Department of Mechanical Engineering, University of California, Riverside, CA, 92521, USA

[3] Department of Physics and Astronomy, University of California, Riverside, CA, 92521, USA

*Corresponding author

**Abstract:** We present a synthesis/processing method for fabricating ferrimagnetic insulator (Bi-doped yttrium iron garnet) thin films with tunable magnetic anisotropy. Since the desired magnetic properties rely on controllable thickness and successful doping, we pay attention to the entire synthesis/processing procedure (nanopowder synthesis, nanocrystalline target preparation and pulsed laser deposition (PLD)). Atomically flat films were deposited by PLD on (111)-orientated yttrium aluminum garnet. We show a significant enhancement of perpendicular anisotropy in the films, caused by strain-induced anisotropy. In addition, the perpendicular anisotropy is tunable by decreasing the film thickness and overwhelms the shape anisotropy at a critical thickness of 3.5 nm.

High quality oxide thin films are the basis for a wide range of optical and magnetic devices. Often the films must be doped to obtain the desired functionality, causing intense work on doped thin film growth by a variety of methods, including molecular beam epitaxy, sputtering and pulsed laser deposition (PLD). The popularity of PLD has increased recently because it has relatively low complexity but is versatile and capable of yielding high quality epitaxial films. The fundamental challenge in producing doped films is transferring the materials from target to substrate with the desired composition. Two different doping approaches are available: 1) Using a doped target with desired constituent elements and 2) Using multiple targets with different constituents as sources. Both have been successful, but the latter so-called combinatorial method requires more complex instrumentation [1] and since the cost and complexity of procedures is often the most important factor in the proliferation of materials in applications, the former approach is more desirable.

It is understandable that the PLD target's characteristics (composition, microstructure, density) should affect film quality, however there is relatively little work devoted to the processing science of target preparation. There has been some work in this regard in un-doped films. Jakeš *et al*[2] used a modified sol-gel (Pechini) technique to make precursors and simultaneously densified and reacted (solid-state reaction) the powders to make lithium niobate targets. They found that thin films prepared from their polycrystalline targets were smoother than those prepared from commercial single crystal targets, suggesting that grain boundaries plays a positive role in film growth. Similarly, Kim *et al*[3] examined the effect of target density on PLD growth of polycrystalline $TiO_2$ thin films, finding that higher density targets yielded smoother films. In work where successful doping was crucial, Sharma *et al*[4] demonstrated ferromagnetism in Mg doped ZnO thin films. The films were grown by PLD from solid-state reacted sintered pellets; Attention to detail in target preparation was important, since too high a sintering temperature resulted in loss of ferromagnetism. In contrast, to these previous studies, we ensure proper doping and phase homogeneity throughout the procedure *i.e.* from powder to bulk target to film. To obtain high quality material we opted for a chemical route for synthesis of fine, nanocrystalline and homogeneously doped powder and then densified the powders to produce a dense PLD target without causing excessive grain growth using Current Activated Pressure Assisted Densification (CAPAD).[5] We then use this homogenous, highly dense nanocrystalline target in PLD deposition.

The ferrimagnetic insulator, yttrium iron garnet, $Y_3Fe_5O_{12}$ (YIG) has drawn considerable attention for use in magneto-optical[6,7] and insulator-based spintronic devices.[8–12] For spintronics, bilayer or multilayer structures containing YIG and a conducting material are often used and such structures require smooth films and high quality interfaces. A common approach is to grow smooth epitaxial thin films on other garnet substrates (often yttrium aluminum garnet (YAG) or gadolinium gallium garnet (GGG)) by PLD. Due to lattice constant match, this results in the easy magnetic direction in the plane of film dictated by the shape anisotropy. Recently we have shown the well-controlled growth of atomically flat YIG films that display in-plane magnetization.[13] For many fundamental studies as well as device applications, it is beneficial to have the magnetization out-of-plane *i.e.* induced perpendicular magnetic anisotropy (PMA). For example, in heterostructures of magnetic insulators and graphene, the exchange interaction provided by the magnetic insulator with PMA is expected to modify the electronic band structure of graphene which consequently causes new physical phenomena such as the quantized anomalous Hall effect.[14] Hence, controlling and tuning magnetic anisotropy has been a goal for various applications.[10,15] In this work, we report a synthesis and processing procedure to engineer magnetic anisotropy in YIG based films using the dual approach of strain induced by doping and substrate thermal expansion mismatch. We chose the heavily doped $Bi_{1.5}Y_{1.5}Fe_5O_{12}$ stoichiometry to achieve this goal, for reasons discussed below.

A polymeric solution based, organic/inorganic entrapment route was employed to synthesize Bi substituted YIG powders, $Bi_{1.5}Y_{1.5}Fe_5O_{12}$. This route has been reported previously for other oxide based systems.[16,17] Commercially available yttrium nitrate hexahydrate, $Y(NO_3)_3 6H_2O$, iron nitrate nonahydrate, $Fe(NO_3)_3 9H_2O$ and bismuth nitrate pentahydrate, $Bi(NO_3)_3 5H_2O$ all from Alfa Aesar® were the precursor source of the cations. Polymeric solution contains 5 wt% of 80% hydrolyzed Polyvinyl alcohol (PVA, Sigma-Aldrich®) dissolved in ultra-high pure water by stirring for 24 h at room temperature. Stoichiometric amounts of chemical precursors were also dissolved and stirred for 24 h before the addition of the polymeric solution. The ratio of nitrate precursors to the PVA solution was chosen in such a way that there were 4 times more positively charged valences from the cations than negatively charged functional end groups of the of PVA, (i.e., −OH groups). This ensures that there were more cations in the solution than the hydroxyl functional groups of the polymer with which they could chemically bond. Then, the precursor solutions were mixed with the polymeric solution and a few drops of nitric acid, ($HNO_3$), were added to the solution to ensure that the pH of the solution was maintained around 0.25 to prevent gelation or precipitation during the reaction. Continuous stirring resulted in complete dissolution of the precursors and the clear solution was then heated on a hot plate (~300 °C) with continuous stirring until the water of the solution formed a thick dark brown resin. The aerated resin formed was then vacuum dried at 125 °C for 24 h. The vacuum drying process resulted in a brownish dried crisp foam, which was first ground using an agate mortar and pestle and heat treated in air at 700 °C for 30 minutes. We used thermal analysis (DSC/TGA, TA Instruments) with a heating rate of 20 °C/min in flowing air to understand the decomposition temperature regime, weight loss, and crystallization temperature. Removing all the volatile matter is of paramount importance to obtain a dense target as well as to employ it in a high vacuum chamber for PLD deposition. The TGA curve of the as-synthesized resin shows no weight loss observed above 700 °C indicating that all volatile material is removed by that temperature (data not shown here). The effect of annealing temperature on phase formation is shown in **Fig 1a.** X-ray diffractometry (XRD) was performed with Cu Kα radiation (PANalytical Empyrean diffractometer) at the room temperature using a step size of 0.013 and 30.60 s per step. XRD analysis of the powders calcined at 700 °C for 30 minutes indicates primarily Bi substituted perovskite phase (Bi:YIP) Bi substituted garnet as a secondary phase (Bi:YIG). The formation of YIP phases is commonly observed during YIG synthesis due to its low energy of formation. Upon further heat treatment, the powders completely convert into the desired garnet structure (Bi: YIG) at temperatures of 800 °C and higher with no remanence of perovskite structure (Bi:YIP). There is also evidence of minor BiFeO$_3$ present at 800 and 900 ºC which is not observed at 700 ºC. This is likely caused by these temperatures being near to, or higher than the melting point of Bi$_2$O$_3$ (817 ºC) causing high reactivity of Bi$_2$O$_3$ with excess Fe$_2$O$_3$ associated with the formation of YIP. The advantage of working with chemical precursor route is the excellent mixing of the starting precursors and high purity of the resultant oxides.[18,19] Good mixing of the starting materials ensures low processing temperature compared to conventional routes, which leads to fine crystallite size. This is visible in **Fig. 1b** showing Scanning Electron Microscopy (SEM, FEI NNS450) micrographs of powders calcined at 700 °C for 30 min. The porous microstructure of the powder is typical of steric entrapment synthesis method and facilitates the grinding process.[17] Low magnification SEM (images are not shown here) reveal that the particles are irregular in shape after initial grinding ranging from 0.1 to 30 μm with majority ~0.5 μm. However, a higher magnification SEM micrograph of the particles revealing nano-scale features ranging from 54 to 190 nm can be seen in the inset in **Fig. 1b**.

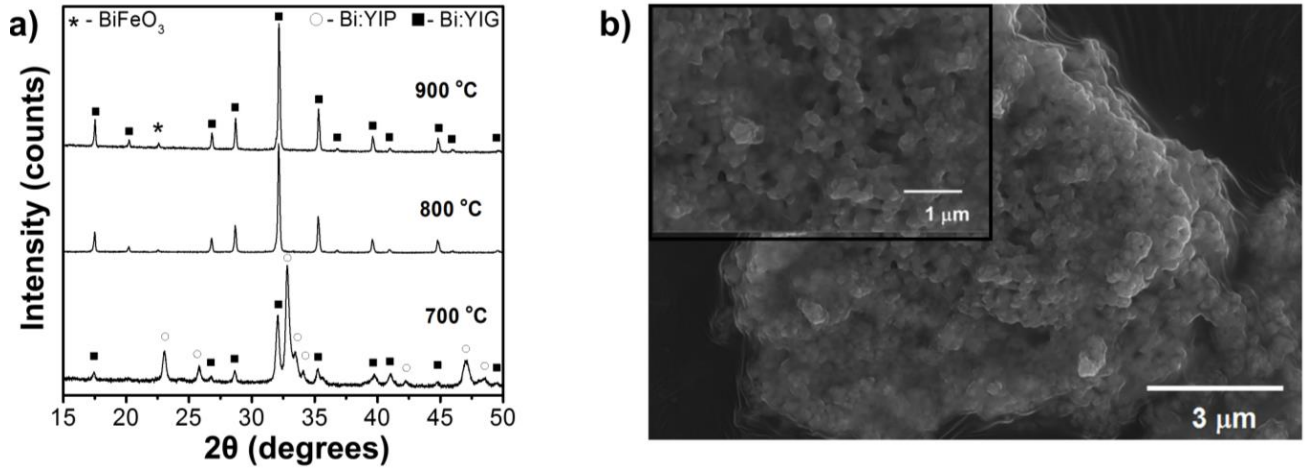

**Figure 1.** Characterization of nanocrystalline Bi:YIG powder produced using the organic/inorganic entrapment route. a) XRD of the powders heat-treated at various temperature for 30 minutes in air. b) SEM images of the powders calcined at 700 °C for 30 minutes. The inset is a higher magnification image of a powder particle revealing nanometer size grains.

Powders calcined at 700 °C were chosen for densification since they offer a good compromise between desired phases and grain size. The powders were then further grinded in order to reduce the particle sizes and to increase the specific surface area in preparation for bulk target processing. A graphite die of 19 mm diameter, with inside cavity wrapped with thin graphite foil was employed to consolidate the crushed powders. Two thin graphite spacers were placed between the powder compact and the top and bottom plungers. The assembly was loaded in to a custom built, CAPAD setup,[5] also known as Spark Plasma Sintering (SPS), 105 MPa pressure was uniaxially applied for 1 minute. The pressure was maintained while the assembly was heated to 700 °C at the rate of 150 °C/min, held for 5 min and cooled down to the room temperature at the same rate. This processing technique takes advantage of benefits of simultaneous application of electric current and pressure which allows for dramatically decreased processing time and temperatures making it possible to retain the nanostructure.[20]

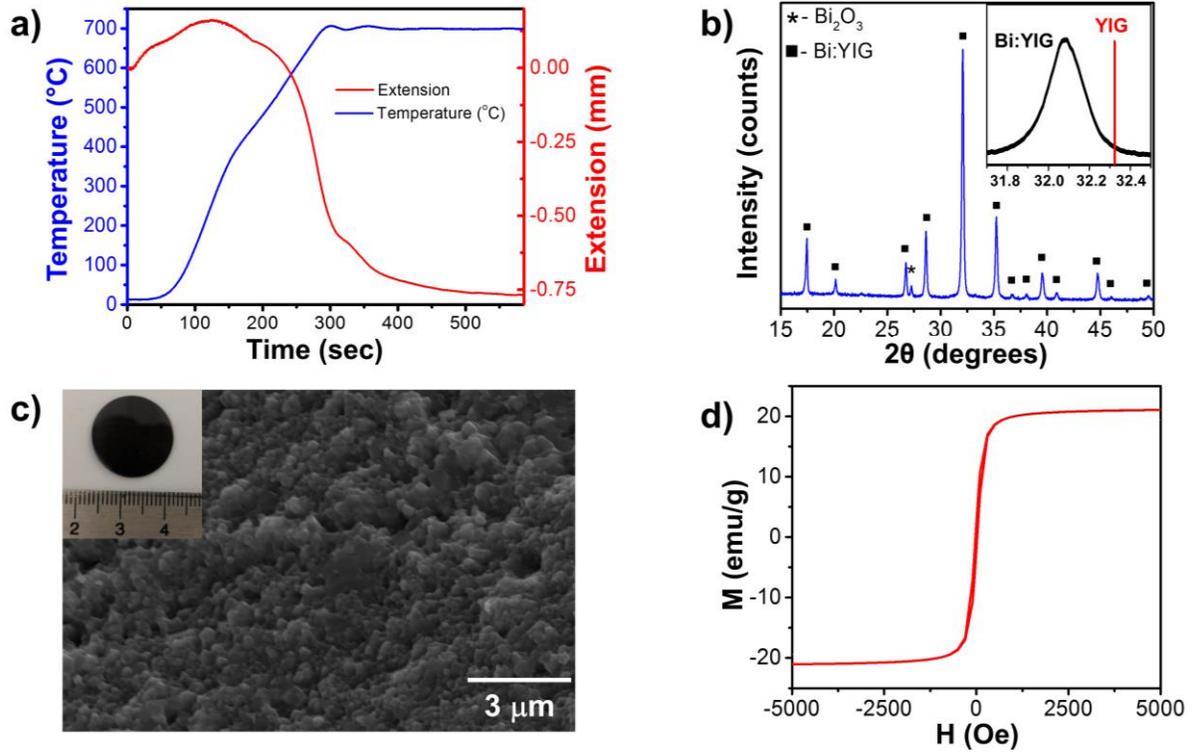

**Figure 2.** Characterization of nanocrystalline Bi:YIG dense targets processed using CAPAD. a) Response of the powder compact during CAPAD processing under 105 MPa at the rate of 150 °C/min. b) XRD of densified specimen (Bi:YIG target) shows high homogeneity after the densification. The inset shows a magnified maximum intensity peak with a corresponding reference YIG peak (red color). c) SEM micrograph of fractured surface of the bulk sample revealing grain size ranging from 80 to 280 nm, with an average of 186 nm. The inset shows a picture of the target. d) Hysteresis loop of the dense target showing soft magnetic behavior.

**Fig. 2a** depicts the various stages of the consolidation involved during CAPAD after removing the equipment compliance under 105 MPa. In stage I, the compacted powders thermally expand due to the heat treatment involved. During stage II, when the die temperature is ~275 °C, densification starts to overcome the thermal expansion, resulting in major shrinkage. The sample continues densifying through the final temperature (700 °C) until it reaches an asymptote at the end of the final stage III. In general, conventional processing of Bi:YIG system over a range of Bi concentration on YIG requires either high temperature (above 900 to 1400 °C), multiple heat treatments and holding for longer time ranging from 1 to 24 h [21–23]. By contrast, in the present work, the high heating rate combined with the applied pressure allowed full densification in less than 600 s at 700 °C.

The XRD pattern of the CAPAD processed target (**Fig. 2b**) indicates the complete phase transformation to Bi:YIG (no evidence of Bi:YIP or $BiFeO_3$ phases) during consolidation with minor $Bi_2O_3$ phase segregation. The absolute density of the bulk, dense material measured is 6.15 g/cm$^3$ which corresponds to 97.3 % of the theoretical density. The theoretical density of 6.32 g/cm$^3$ is obtained using lattice parameters values measured on the powder samples heat-treated at 900 °C for 30 minutes. Parameters from powder sample rather than bulk sample were used to avoid the influence of micro-strain, which could possibly deviate the lattice parameter values by peak broadening. The inset on **Fig. 2b** shows the major peak of the consolidated sample in comparison with the standard YIG reference peak (red line); it clearly reveals that the substitution of $Bi^{+3}$ on $Y^{+3}$ site resulted in an increased lattice constant and shifted the peak toward lower angle.

Microstructures of fractured surface of the CAPAD processed bulk sample are shown in **Fig 2c**, revealing a fine grained microstructure with some fine pores ($\leq 1$ µm), the average grain size is 186 nm (ranging from 80 to 280 nm). The inset in the **Fig. 2c** is a picture of the target (scale in cm). BSE images of polished surfaces (not shown here) showed no noticeable phase segregation. We attribute the fine homogenous microstructure to the optimized processing conditions, especially to the low densification temperature. Using higher temperatures and conventional solid-state processing would surely cause significant grain growth and possibly a heterogeneous microstructure since liquid phase sintering is the active densification mechanism (melting temperature of $Bi_2O_3$ is 817 °C).[22,24]

The magnetic properties of the CAPAD processed dense specimen were measured using a vibrating sample magnetometer (VSM) at room temperature with an applied magnetic field up to 10 kOe. **Figure 2d** shows a hysteresis curve of the Bi:YIG, indicating a soft magnetic behavior as expected. The saturation magnetization ($M_s$) is 21.1 emu/g, remanence ($M_r$) is 2.9 emu/g and an intrinsic coercive force ($H_c$) is 54 Oe. These values are consistent with previously reported values for Bi:YIG system with similar compositions.[18,23]

High quality, ultraflat Bi:YIG thin films were grown on (111) oriented YAG substrates by PLD system using the CAPAD processed Bi:YIG targets. The substrates were first cleaned with acetone followed by isopropyl alcohol and deionized water. Prior to deposition, the substrates were annealed at moderate temperature ~200 °C overnight in high vacuum at the level of $10^{-7}$ torr. After gradually increasing the substrate temperature to about 800 °C and oxygen pressure with 12 wt% ozone to 1.5 mTorr, the KrF excimer laser pulses of 248 nm in wavelength with power of 150 mJ struck the target at a repetition frequency of 1 Hz. The deposition rate is estimated to be 1 Å/min. After deposition, the film was annealed under the same oxygen pressure for an half hour in order to enhance the crystalline structure before slowly decreasing the heater temperature and oxygen pressure. AFM analysis (**Figure 3a**) indicates atomically flat films with low roughness (~0.5 nm) and no pin holes were found on films. XRD of the 7 nm thick Bi:YIG thin film grown on YAG substrate shows the major YAG single crystal peak as expected, in addition to the Bi:YIG (444) peak (**Figure 3b**). In addition, the analysis reveals that the films contain no secondary phases (the peak at ~51° is due to the YAG substrate).

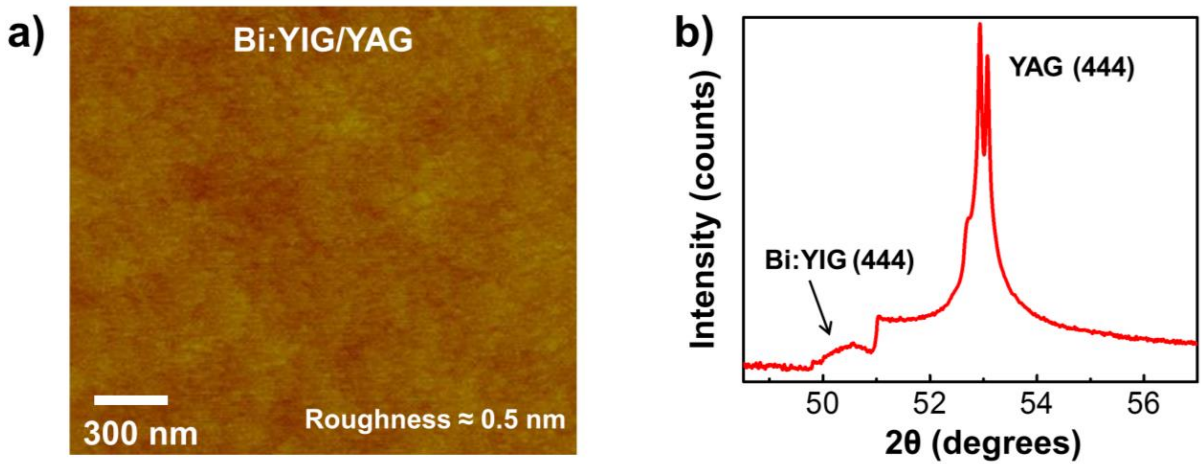

**Figure 3.** Structural characterization of Bi:YIG films a) AFM surface profile for $Bi_{1.5}Y_{1.5}Fe_3O_{12}$ YIG film with a root-mean-square roughness about 0.5 nm across a 2 μm × 2 μm scan area. b) XRD pattern of the same Bi doped YIG films grown on YAG (111).

In YIG thin films grown on YAG substrate, magnetic anisotropy $K_u$ is dominated by the shape anisotropy, resulting in the easy axis directed in-plane, as seen in the inset of **Fig 4a**. One could tune the magnetic anisotropy by introducing lattice constant mismatch induced magnetostriction.[15,25] Along the <111> orientation of cubic crystals, the magnetostriction effect induced effective perpendicular anisotropy field is described as $H_\perp \sim \frac{-3\lambda_{111}\sigma_\parallel}{M_s}$,[26] where the $\lambda_{111}$, $\sigma_\parallel$, and $M_s$ are the magnetostriction constant, in-plane stress, and saturation magnetization, respectively. The negative value of $\lambda_{111}$ (-4.45 × 10$^{-6}$) in garnet films requires a tensile stress ($\sigma_\parallel >$ 0) in order to produce a positive perpendicular anisotropy field, needed for PMA. The stress is given by $\sigma_{\parallel} = -\frac{Y}{2\mu}\frac{a_\perp - a_o}{a_o}$ where $a_\perp$ is the lattice constant perpendicular to the film, $Y$ is Young's modulus, $\mu$ is Poisson's ratio and $a_o$ is the unstrained lattice parameter of the film. In $Bi_xY_{1-x}Fe_3O_{12}$ films,[27] in the regime where the Bi doping is light, it is found that an increasing Bi content $x$, results in a larger perpendicular lattice constant based on a homogeneous elastic compression because of the small lattice mismatch. However, in the regime of heavy doping of Bi ($x \geq 1.5$), the stress is accommodated by dislocations especially at elevated growth temperatures. Due to the thermal expansion difference between the film and substrate as the film is cooled back to room temperature, stress is built up at the interface at room temperature (region 2 in **ref 27**). The strained lattice constant, in such heavily Bi doped YIG is $a_\perp = a_o - \frac{2\mu}{1-\mu}(\alpha_f a_o - \alpha_s a_s)\Delta T$, where $\alpha_f$ and $\alpha_s$ are thermal expansion coefficient for the films and substrates, respectively and $\Delta T$ is the difference between growth and room temperature.[27] In this case, the films are subjected to a tensile strain, owing to the larger coefficient for the $Bi_xY_{1-x}Fe_3O_{12}$ films ($\alpha_f > (10.4 + 0.83x) \times 10^{-6}/°C$) than that of substrates ($\alpha_f(YAG) = 7.2 \times 10^{-6}/°C$). Therefore, in the heavy Bi-doping regime such as the $x = 1.5$ stoichiometry chosen here, it is promising to generate PMA in Bi:YIG films grown on YAG, thanks to the larger tensile strain.

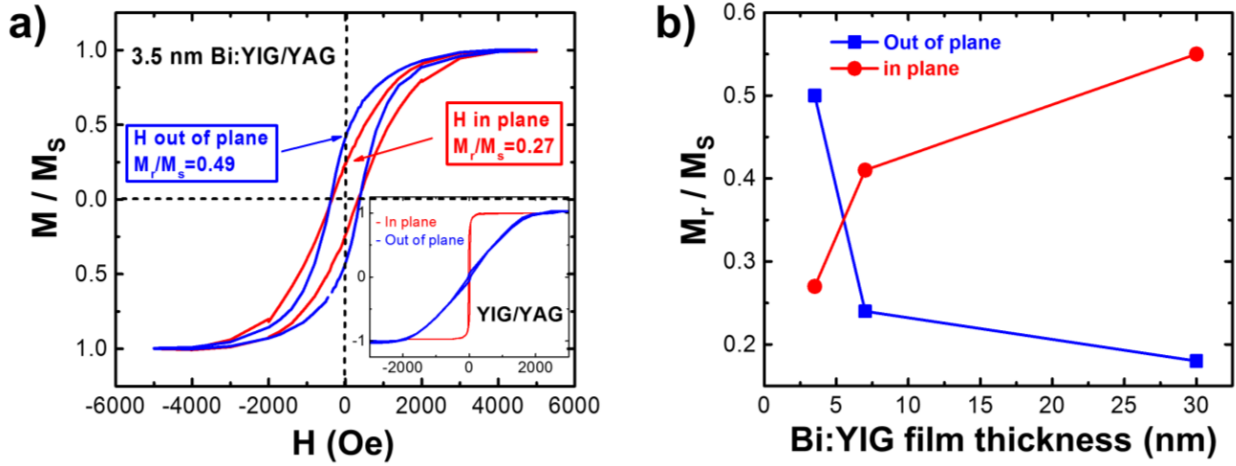

**Figure 4.** a) Normalized magnetic hysteresis loops for both field out-of-plane and in-plane geometries of 3.5 nm $Bi_{1.5}Y_{1.5}Fe_3O_{12}$ film grown on (111) oriented YAG after removing the diamagnetic background of the substrates. Inset: Normalized magnetic hysteresis loops for both out-of-plane and in-plane geometries of undoped YIG films grown on (111) oriented YAG after removing the diamagnetic background of the substrates showing a strong in-plane anisotropy. b) The thickness dependence of the squareness for $Bi_{1.5}Y_{1.5}Fe_3O_{12}$ films on YAG for both in-plane and out-of-plane indicating the enhancement of perpendicular magnetic anisotropy for thinner films.

Magnetic property of the as-grown thin films was then measured using VSM and a Superconducting Quantum Interference Device (SQUID). **Figure 4a** shows the normalized magnetic hysteresis loop of 3.5 nm $Bi_{1.5}Y_{1.5}Fe_3O_{12}$ film grown on (111) oriented YAG for both field out of plane and in-plane geometries after removing the diamagnetic background of the substrates. In contrast to similarly grown YIG films on YAG (inset of **Fig. 4a**), the out-of-plane curve is comparable to the in-plane curve, clearly indicating an increase in PMA of the Bi:YIG film. We attribute the increase in PMA to interfacial strain caused by difference in thermal expansion coefficients as discussed above. It is also a possibility that a Dzyaloshinskii-Moriya (DM) interaction could contribute. In thin films heterostructures, DM interaction usually arises from the inversion symmetry broken at the interface. Both the DMI and PMA, to first approximation are proportional to the strength of spin orbit coupling (SOC). In another set of experiments (not shown here) we have deposited Bi:YIG films on other GGG, yielding similar results. In addition, we have deposited other rare earth iron garnets ($Tm_3Fe_5O_{12}$) on the combination of garnet substrates (SGGG and GGG) which also show strain tunable PMA[28]. Similar findings amongst different rare earth garnets on different substrates, strongly suggest that interfacial strain is a dominant over possible DM interaction. In order to further confirm that increase in PMA is an interfacially induced effect, we conducted a thickness dependence study. **Figure 4b** shows the Bi:YIG thickness dependence of the squareness of magnetic hysteresis loops defined as the ratio of remanence $M_r$ over saturation magnetization $M_s$ for both in-plane and out of plane. A strong enhancement of the squareness for out of plane and suppression of squareness for in-plane when Bi:YIG films become thinner suggests the increase of PMA is caused by the interfacial strain. The perpendicular magnetic anisotropy increases with a decrease in film thickness and overwhelms the shape anisotropy at a thickness of 3.5 nm.

In summary we have presented an efficient method for the production of thin Bi:YIG films grown from well doped nanocrystalline targets which were made form high quality nanocrystalline powders. The films have strain induced PMA from a combination of the Bi doping and thermal expansion mismatch. These results should be useful for the continued development of ferrimagnetic insulator based spintronic studies and devices where PMA is important.

**Acknowledgement**

This work was supported as part of the Spins and Heat in Nanoscale Electronic Systems (SHINES), an Energy Frontier Research Center funded by the U.S. Department of Energy, Office of Science, Basic Energy Sciences (BES) under award # SC0012670.